\documentclass{aa}

\usepackage{graphicx}
\usepackage{microtype}
\usepackage{xcolor}
\usepackage{txfonts}
\usepackage{hyperref}
\usepackage{listings}
\usepackage[normalem]{ulem}

\begin{document}
    
    \title{SNGuess: A method for the selection of young \\ extragalactic transients}

    
    \author{
    N. Miranda\inst{\ref{inst1}} \and
    J.C. Freytag\inst{\ref{inst1}} \and
    J. Nordin\inst{\ref{inst2}} \and
    R. Biswas\inst{\ref{inst5}} \and
    V. Brinnel\inst{\ref{inst2}} \and
    C. Fremling\inst{\ref{inst3}} \and
    M. Kowalski\inst{\ref{inst2}}\inst{\ref{inst6}} \and
    A. Mahabal\inst{\ref{inst3}}\inst{\ref{inst4}} \and
    S. Reusch\inst{\ref{inst2}}\inst{\ref{inst6}} \and
    J. van Santen\inst{\ref{inst6}}
    }

    \institute{
            Institut f\"{u}r Informatik, Humboldt-Universit\"{a}t zu Berlin, Rudower Chaussee 25, 12489 Berlin, Germany\label{inst1}
            \and
            Institute of Physics, Humboldt-Universit\"{a}t zu Berlin, Newtonstr. 15, 12489 Berlin, Germany\label{inst2}
            \and
            The Oskar Klein Centre, Department of Physics, Stockholm University, AlbaNova, SE-106 91 Stockholm, Sweden\label{inst5}
            \and
            Division of Physics, Mathematics, and Astronomy, California Institute of Technology, Pasadena, CA 91125, USA\label{inst3}
            \and
            Deutsches Elektronen-Synchrotron, D-15735 Zeuthen, Germany\label{inst6}
            \and
            Center for Data Driven Discovery, California Institute of Technology, Pasadena, CA 91125, USA\label{inst4}
            }

    \date{Received -; accepted -}

 
    \abstract
   {With a rapidly rising number of transients detected in astronomy, classification methods based on machine learning are increasingly being employed. Their goals are typically to obtain a definitive classification of transients, and  for
good performance they usually require the presence of a large set of observations.
   However, well-designed, targeted models can reach their classification goals with fewer computing resources.}
   {The aim of this study is to assist in the observational astronomy task of deciding whether a newly detected transient warrants follow-up observations.}
   {This paper presents \texttt{SNGuess}, a model designed to find young extragalactic nearby transients with high purity. \texttt{SNGuess} works with a set of features that can be efficiently calculated from astronomical alert data. Some of these features are static and associated with the alert metadata, while others must be calculated from the photometric observations contained in the alert. Most of the features are simple enough to be obtained or to be calculated already at the early stages in the lifetime of a transient after its detection. We calculate these features for a set of labeled public alert data obtained over a time span of $15$ months from the \emph{Zwicky Transient Facility (ZTF)}. The core model of \texttt{SNGuess} consists of an ensemble of decision trees, which are trained via gradient boosting.}
   {
   Approximately $88\%$ of the candidates suggested by \texttt{SNGuess} from a set of alerts from ZTF spanning from April 2020 to August 2021 were found to be true relevant supernovae (SNe).
   For alerts with bright detections, this number ranges between $92\%$ and $98\%$.
   Since April 2020, transients identified by \texttt{SNGuess} as potential young SNe in the ZTF alert stream are being published to the Transient Name Server (TNS) under the \texttt{AMPEL\_ZTF\_NEW} group identifier. \texttt{SNGuess} scores for any transient observed by ZTF can be accessed via a web service\thanks{\url{https://ampel.zeuthen.desy.de/api/live/docs}}. The source code of \texttt{SNGuess} is publicly available\thanks{\url{https://github.com/nmiranda/SNGuess}}.}
   {
   \texttt{SNGuess} is a lightweight, portable, and easily re-trainable model that can effectively suggest transients for follow-up. These properties make it a useful tool for
   optimizing
   follow-up observation strategies and for assisting humans in the process of selecting candidate  transients.
   }

             

   \maketitle
%

\section{Introduction}
\label{sec:intro}

The study of transient astrophysical events has made it possible to understand explosive phenomena not accessible in terrestrial laboratories, and to map out the evolution of the Universe.
Type Ia supernovae (SNeIa) are particularly important in this context. 
Their use as standardizable candles has allowed astrophysicists for already more than two decades to accurately measure the distance of remote regions of the Universe, and they provide the opportunity to gain insight into the mechanisms of stellar death \citep{riess_observational_1998}.

A new generation of astronomical surveys, including the currently operating Zwicky Transient Facility (ZTF; \citealt{bellm_zwicky_2018}), the All-Sky Automated Survey for SuperNovae (ASAS-SN; \citealt{kochanek_all-sky_2017}), the Asteroid Terrestrial-impact Last Alert System (ATLAS; \citealt{tonry_atlas_2018}), the Dark Energy Survey (DES; \citealt{abbott_first_2019}), the Panoramic Survey Telescope and Rapid Response System (Pan-STARRS; \citealt{kaiser_pan-starrs_2002}), and the upcoming Legacy Survey of Space and Time (LSST; \citealt{ivezic_lsst_2019}) conducted on the Vera C. Rubin Observatory, give the community unprecedented real-time access to observations of these events.

These facilities perform automated photometric observations of many sources in large regions of the sky, and then distribute the data to the community via alert streams. Subsequently, the observational sources that are deemed to be of particular interest are inspected in more detail by photometric and/or spectroscopic follow-up observations. Follow-up observations allow researchers to precisely identify and characterize astrophysical phenomena.

Spectroscopic resources are limited, especially when compared to the large number of astronomical transients that are photometrically detected and distributed as alerts. The fraction of photometric candidates that are spectroscopically classified is therefore rapidly decreasing with improved photometric surveys. Explosive phenomena such as SNe and other fast transients often have a short lifetime (usually weeks or months) and their behavior evolves in a matter of days, or even hours.

Early detection and follow-up of explosive transients is particularly relevant in the study of SNeIa. 
It is widely agreed upon in the community that SNe of this kind are the result of thermonuclear explosions that originate in the interaction between two progenitor stars: a white dwarf and a companion. 
However, there remain open questions regarding the exact nature of both these progenitors and of the mechanism of the resulting explosion \citep{maoz_observational_2014}.

Many of these questions can only be studied from observations that take place early in the lifetime of the explosive transient phenomenon.
For instance, early photometric observations of a SNeIa can constrain the radii of the possible progenitor white dwarf star and its companion \citep{nugent_supernova_2011}. 
The speed of the increase in luminosity after the explosion and its duration are a useful probe of the inner and surrounding distribution of material at the core of the white dwarf \citep{dessart_constraints_2014}. 
Also, strong emission at certain wavelengths early after explosion may indicate particular interactions between the ejected material and the companion star \citep{kasen_seeing_2010}.

Having greater insight into the aforementioned physical phenomena would vastly improve the ability of astrophysicists to calibrate SNeIa for their use as standardizable candles, and allow a better understanding of the evolution of large-scale structures of the Universe.

In general, automated methods for the classification of astronomical transients based on photometric data are expected to become critical tools both for parsing the large alert streams in real-time streams and for understanding the full, final observed sample. In recent years, photometric classification challenges such as PLAsTiCC \citep{hlozek_results_2020} and SNPhotCC \citep{kessler_results_2010} have succeeded in bringing together the expertise of both the astronomical and machine learning communities in the development of new state-of-the-art tools.

Methods commonly used for time-series classification broadly fall into two groups: feature-based and data-driven (or nonparametric).
Feature-based classifiers rely on inferring qualities and substructures from the measurements, according to previously defined (or engineered) functions, and then using these as a representation of the time series.
Most features are statistical and structural metrics that can be calculated over a collection of measurements in time \citep{schafer_teaser_2020}.
In time-domain astronomy, feature functions are selected for their application based on their ability to reflect specific characteristics of time-variable astrophysical phenomena.
We can find some examples of feature-based classifiers in ALeRCE \citep{sanchez-saez_alert_2021}, \texttt{Avocado} \citep{boone_avocado_2019}, and \cite{dai_photometric_2018}.

On the other hand, data-driven methods operate directly on the measurements of the time series, and do not depend on explicitly defined features.
Thus, in this case it is the process of building these features that is explicitly defined, and this is integrated into the automated learning method itself.
We can find some examples of data-driven methods in \cite{charnock_deep_2017}, \cite{mahabal_deep-learnt_2017}, \texttt{PELICAN} \citep{pasquet_pelican_2019}, \texttt{RAPID} \citep{muthukrishna_rapid_2019}, \texttt{SCONE} \citep{qu_photometric_2022}, \texttt{snmachine} \citep{alves_considerations_2022}, \texttt{SuperNNova} \citep{moller_supernnova_2020}, and \texttt{SuperRAENN} \citep{villar_superraenn_2020}.

Another way to categorize these methods is according to which phase of the transient explosion they would focus on in order to classify according to transient type. Methods such as the one developed by \cite{dai_photometric_2018} take light curves that display the complete life cycle of the explosive transient  as input, while those like \textit{SCONE} can take partial time series as input \citep{qu_photometric_2022}. Others such as \texttt{RAPID} are designed for both cases \citep{muthukrishna_rapid_2019}.

Even though they are successful in classifying transient candidates according to photometric data, these models are built on complex architectures (especially in the case of deep neural networks), which makes it hard for researchers to grasp the underlying logic of their decisions. 
Furthermore, most of these models require several observations in order to achieve their high classification performance, making them less suited to the identification of young transients. 

The astrophysical transient research community has already recognized the importance of follow-up observation strategy planning, given the limited resources and the difficulty of obtaining reliable labels from which supervised classification algorithms can learn. 
To this end, active learning strategies have recently been used to design training sets for machine learning classifiers, in the context of the peculiar data environment of astronomical transient detection (\citealt{ishida_optimizing_2019}; \citealt{kennamer_active_2020}; \citealt{leoni_fink_2022}; \citealt{carrick_optimizing_2021}).

Some of the classification algorithms previously mentioned, such as RAPID and \texttt{snmachine}, have the specific use case of performing fine-grained classification between SN subtypes. Therefore, they were trained and tested using explosive transient data only.
It is assumed, in these cases, that other types of transients, such as moving objects, cataclysmic variable stars (CV), and active galactic nuclei (AGN) can easily be removed by comparison with historical observations obtained far back in time.
However, this is hard to do for surveys that are in their starting phase.

For some surveys, once nontransient sources of variability such as AGN and variable stars are discarded, the resources available are sufficient to classify nearly all explosive candidates that reach a certain peak magnitude.
For instance, in the case of ZTF, the value of this peak magnitude lies around 18.
In the case of LSST, the Time-Domain Extragalactic Survey (TiDES; \citealt{swann_4most_2019}) is expected to follow up all explosive transients detected with magnitudes $r_\mathrm{AB} \lesssim 22.5$ at peak.

For these reasons, here we study the effectiveness of an alternative method that is able to provide an informed guess of whether a particular variable candidate will become an explosive transient that is relevant for follow-up. This guess should take place at early observation times and without the need for many historical observations or redshift-related catalog information.

This article is structured as follows. In Section \ref{sec:terminology} we introduce the astronomical and classification concepts (the latter in the context of machine learning) used in this text. In Section \ref{sec:motivation} we discuss the motivation and goals taken into consideration when designing and implementing \texttt{SNGuess}. Sections \ref{sec:feature_extraction} to \ref{sec:training} describe the main steps followed in order to train the model that is at the core of \texttt{SNGuess} (see Fig. \ref{fig:snguess_steps}). Section \ref{sec:model_deployment} gives details of how \texttt{SNGuess} is actively selecting transients and how the results have been made available to the community and Section \ref{sec:testing} evaluates the performance. Finally, in Section \ref{sec:conclusions} these results are summarized and we list important challenges to tackle in forthcoming work.

\section{Terminology}
\label{sec:terminology}

This study makes use of terminology both from astronomy and from computer science; specifically from the domains of data science and machine learning, and of classification tasks (or alternatively, information retrieval).
In this section we proceed to introduce concepts that are  commonly used in transient astronomy and in the classification process\footnote{These are introductions to terminology for a cross-community context. Any reader already familiar with these concepts should skip this section and continue with Sect. \ref{sec:motivation}}.

\subsection{Astronomical concepts}
\label{subsec:astro_conc}

Hereafter, we refer to {"alerts", which are} data packets issued at a certain point in time by an astronomical survey or a highly automated observation facility, such as the ZTF. A {"stream"} or {"alert stream"} is a set of alerts generated and distributed by a given observation facility during an interval of time.

An individual measurement performed by an instrument at the observation facility is called an "{observation}". In optical astronomy, an observation is represented by an image of a region of the sky taken by the camera of the instrument at a particular time.
Point sources of variability in the sky are typically identified from a "difference image"; which is an image that is created through subtracting a newly made observation from a reference image. A reference image is an aggregation of observations of the same region of the sky but generated at previous moments in time. A variable source is identified by the survey every time a point source is recognized in a difference image. A point source in the difference image which exceeds some survey-specific signal-to-noise-ratio (S/N) threshold is called a "{detection}"\footnote{In Rubin Observatory nomenclature, this is called a "{source"} instead of a "{detection}". Readers familiar with that nomenclature should bear this in mind for the remainder of the article.}.

We use the terms "{astronomical source}" or "{astronomical object}" to refer to the astrophysical entity or phenomenon from which a set of detections of a certain point or region in the sky seems to have originated. 
In the context of astronomical alerts, we call these "candidates", to reflect the fact that they may or may not end up being relevant for follow-up observations and subsequent characterization as a particular kind of astrophysical phenomenon.

Alerts are generated by an automated survey each time a detection is made in a difference image.
Observational data of the detection are then complemented with other contextual data into an alert and distributed \citep{bellm_zwicky_2018}.
The alerts may contain metadata related to the observation process itself (this mostly concerns the instrument that was used and the observing conditions) or related to the candidate (such as its location or its proximity to other previously identified sources).

\begin{figure}
\centering
\resizebox{\hsize}{!}{\includegraphics{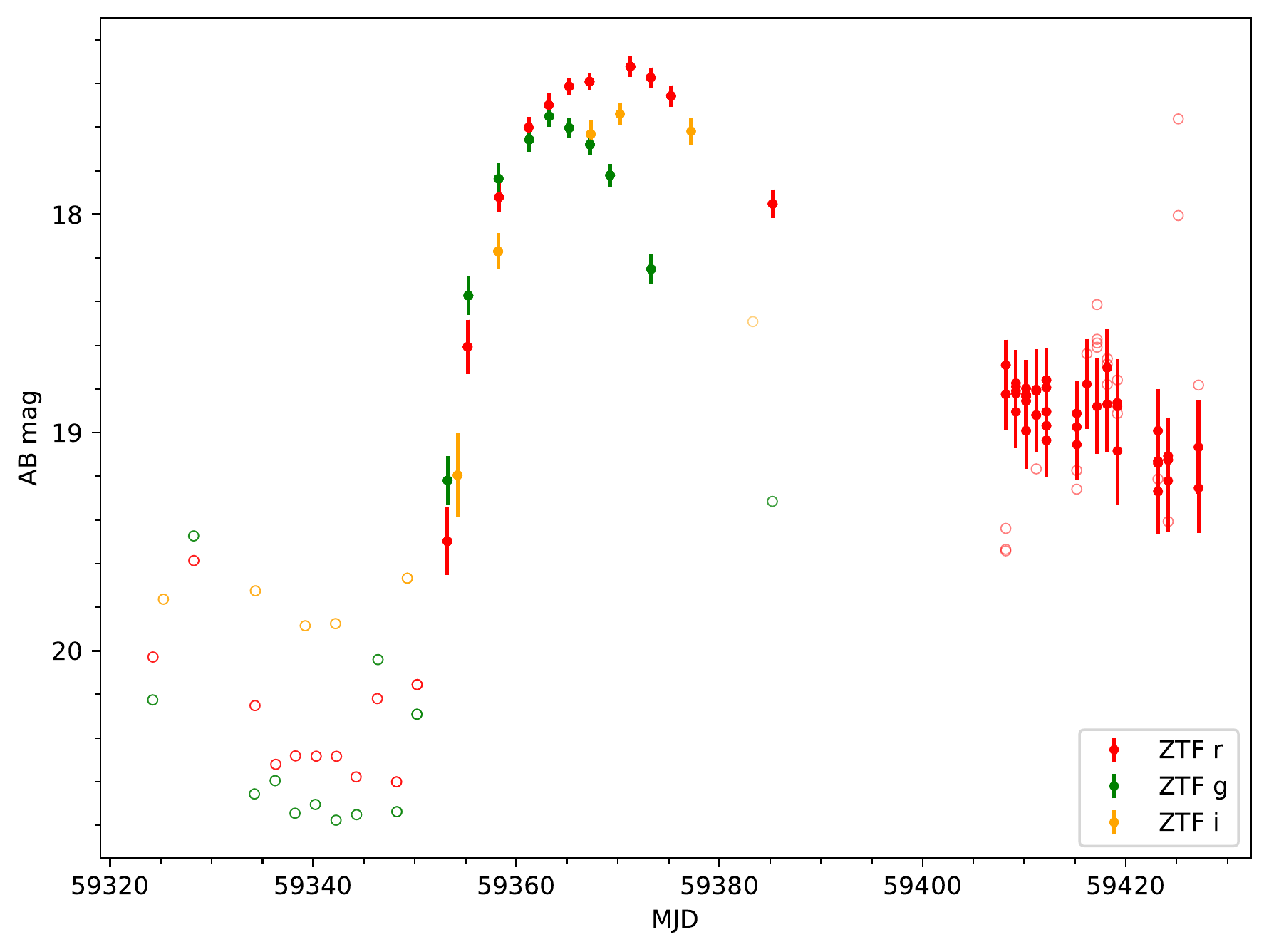}}
\caption{Example of a light curve obtained from a ZTF alert; in this case candidate ID ZTF21abbyhvw. This particular candidate has detections in three bands. The irregular sampling that is characteristic in astronomical observations can be seen. Lightly colored hollow points correspond to nondetections.}
\label{fig:light_curve}
\end{figure}

The nature of the observations contained in the alert will depend on the instrument that was used. For instance, ZTF uses an optical instrument, and therefore its observations are photometric in nature. By "{light curve}" we refer to a representation of a time series as a vector of photometric measurements at a certain band in wavelength and their associated time-stamps. Light curves can have measurements in multiple bands, and they usually have irregular sampling (see Fig. \ref{fig:light_curve}).

If a candidate has a recognizable behavior only during a certain time duration (usually days or weeks) before returning to a "normal" or baseline behavior, we say that the behavior is "{transient}" in nature, or that we observe a {transient} candidate or phenomenon. 
Events, or detections, can be caused by different phenomena, including "proper" transients (e.g., SNe), reoccurring variables (e.g., AGN), moving objects in the Solar System, cosmic rays, or pure noise due to for example misaligned references.

Alerts are represented by binary files containing metadata and contextual information for a single detection or event. This event is usually a change in the position or luminosity of a particular point in the sky \citep{masci_zwicky_2019}. Each alert includes at least a unique identifier and data related to the particular science observation such as, for instance, a cutout image of the observation region \citep{juric_data_2020}. Additionally, in surveys such as ZTF and LSST, alerts include some listing of prior variability connected to the same source. ZTF alerts contain the photometry of any prior detection at the same position during the last $30$ days. 
ZTF alert packages are represented and structured in JSON format. Some of its values are: a unique identifier for the alert, candidate-specific metrics from the image-subtraction process, image-specific metadata, nearby astronomical objects according to different catalogs, and image cut-outs centered on its location.
There is typically a one-to-many relationship between candidates and their alerts. That is, a new alert for a candidate is issued every time a new valid observation is made. Therefore, many alerts are usually produced for a candidate during its transient lifetime.

\subsection{Classification concepts}
\label{subsec:class_concepts}

\begin{table}
\caption{Definitions of some common classification assessment metrics.}
\centering
\begin{tabular}{c c}
    Precision & $\dfrac{\mathrm{tp}}{\mathrm{tp} + \mathrm{fp}}$ \\ 
    Recall & $\dfrac{\mathrm{tp}}{\mathrm{tp} + \mathrm{fn}}$ \\
    F1-score & $\dfrac{2\mathrm{tp}}{2\mathrm{tp} + \mathrm{fp} + \mathrm{fn}}$ \\
    MCC & $\dfrac{ \mathrm{tp} \times \mathrm{tn} - \mathrm{fp} \times \mathrm{fn} }{ \sqrt{(\mathrm{tp} + \mathrm{fp}) (\mathrm{tp} + \mathrm{fn}) (\mathrm{tn} + \mathrm{fp}) (\mathrm{tn} + \mathrm{fn})} }$
\end{tabular}
\tablefoot{Here, \textit{tp} stands for true positive examples, \textit{tn} for true negatives, \textit{fp} for false positives, and \textit{fn} for false negatives.}
\label{table:metrics}
\end{table}

In the context of information retrieval, precision is defined \citep{chicco_advantages_2020} as the fraction of retrieved documents that are relevant (see Table \ref{table:metrics}). Its dual metric, recall, is defined as the fraction of relevant samples that are correctly retrieved. This definition can be directly applied to classification tasks if we interpret relevant and irrelevant samples as positive and negative classes, respectively.
The F1-score, in turn, is defined as the harmonic mean between precision and recall.
The Matthews correlation coefficient (MCC) is a statistical rate that is used as a score for classification tasks. It has the advantage of being more informative and truthful in evaluating binary classifications than the accuracy or the F1-score \citep{chicco_advantages_2020}.

\begin{figure}
\centering
\resizebox{\hsize}{!}{\includegraphics{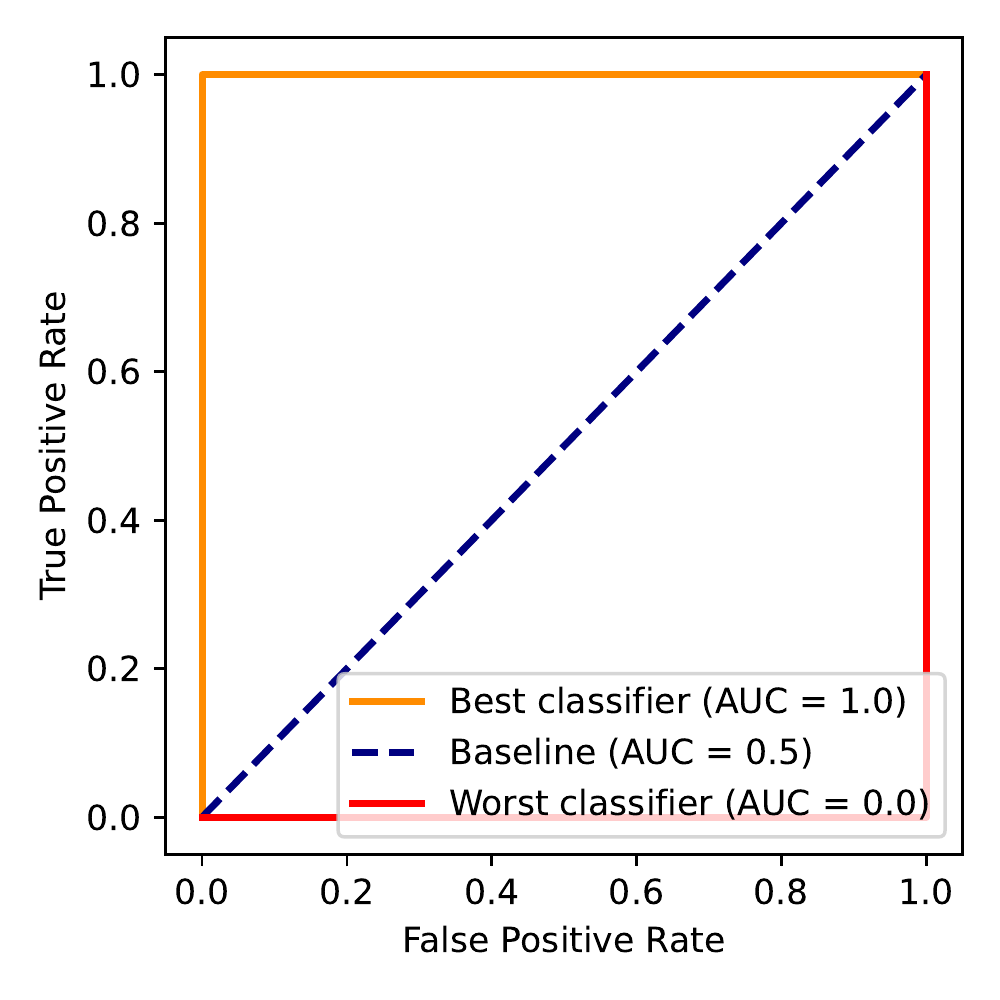}}
\caption{Example of ROC curves with their respective AUC. Here we see the curves that correspond to the performance of the best possible classifier (AUC = 1.0), a perfectly random or baseline classifier (AUC = 0.5), and the worst possible classifier (AUC = 0.0).}
\label{fig:roc_curve}
\end{figure}

The {Receiver Operating Characteristic (ROC)} curve (see Figure \ref{fig:roc_curve}) is the result of plotting true positive rates against false positive rates while gradually changing the threshold value returned by the classification model and used for distinguishing between classes. A ROC curve is plotted in a two-dimensional Cartesian coordinate system, where the horizontal axis represents the false positive rate and the vertical axis represents the true positive rate. Both rates can take values ranging from $0$ to $1$; and therefore, the ROC curve is bounded by the unitary quadrant of the plane.

In a perfect binary classification procedure, if the threshold condition for class distinction is incrementally relaxed (i.e., subsequently more and more examples start being considered as belonging to the positive class), almost all marginal increases in true positive rates occur with no increase in false positive rate. A classification like this will be represented by a curve in the ROC diagram that has the highest possible area under it, and it will have a value very close to $(0,1)$ at some point. This curve will be similar to a Pareto cumulative distribution with an infinite \emph{shape} ($\alpha$) value.

In contrast, if a classification is the worst possible, almost all marginal increases in false positive rate occur with no associated increase in true positive rate. When plotted in a ROC diagram, this curve will be very close to $y=0$ in most of its points, and it will have an area under the curve that is close to zero.

A random binary classification where a given example has the same probability of being assigned to either of the two classes (e.g., flipping a balanced coin each time to decide which class to assign an example) should, if statistically significant, increase the true positive and the false positive rate  by the same marginal amount. This constitutes the baseline performance, and when plotted in the ROC curve it describes a line that corresponds to the identity function (it goes from $(0,0)$ to $(1,1)$).

The ROC curve can be summarized by a single number: the area under the curve (AUC, in this case the ROC AUC). This number is used by itself as a metric to assess classification performance, and its value ranges from $0$ (worst possible performance) to $1$ (best possible performance).

\section{Motivation}
\label{sec:motivation}

The task of classifying light curves has two different goals, depending on whether the potentially observed transient is at an early or late stage of its lifetime.
The scope and the challenges in these two cases are distinct from each other. In the case of late stage classification, the transient has already faded by definition. That is, we collected all observations for the full span of the transient's life, and the goal is to achieve precise classifications, even up to the subtype level. How fine-grained the classification should be depends on the specifics of each science use case. 
When a sufficient number of observations are available, it is trivial to identify Solar System objects, satellites, or other variables simply by looking at their
light curve but not transient sources that are irrelevant for further analysis.

On the other hand, classifying transients at their early stage is fundamentally different: the goal here is to find relevant phenomena as early as possible to the beginning of their lifetime in order to request additional dedicated observations quickly. 
That is, we are interested in detecting quickly developing extragalactic objects in their infant or early phase.
For such objects, we possess a limited number of measurements, and therefore, it is not trivial to distinguish them from irrelevant candidates.
The likelihood of mistaking a transient for another kind of variable object is even higher when the survey is just starting its operations.
This is also the case when there are not many historical observations available overall.
In this context, our main sources of contamination are satellites, Solar System objects, intragalactic objects, and other types of nontransient variable sources, with the possible added presence of observation artifacts.
Furthermore, if we are only focused on following-up young SNe, then distant and old but still bright SNe may be a source of confusion as well.

We investigate if simple but specialized methods could prove successful in terms of achieving early stage classification. In particular, the goal is to develop a model that works as a filter in terms of selecting young SNe as candidates for fast and/or automatic follow-up. In this case, noise will consist mostly of artifacts, nonextragalactic transients, and nontransient variable sources. In our work, the main goal is to reduce the number of false-positive selections by  as much as possible, as the resources for follow-up observations are highly limited. 

The selection of relevant transients for follow-up is a stage dominated by intense human labor. Any method that may assist in this process should be a valuable contribution in improving the efficiency of the whole transient detection and characterization pipeline.

Our selection method, which we call \texttt{SNGuess}, consists of extracting a feature vector from each alert of a set, and then organizing the vectors into a feature matrix. In a next step, a set of labels that indicate the relevance for follow-up of each one of the alert candidate objects is obtained (one label for each one of the feature vectors) and is then used to construct a target vector. The target vector takes Boolean values (\texttt{true} when the alert candidate is relevant, \texttt{false} when it is not relevant). The feature matrix and the target vector provide the input for supervised training of a boosted-trees classification model.

Once the model is trained, vectors of the same feature functions are extracted; this time from a different set of alerts to be analyzed. These vectors are subsequently fed to the trained model in order to obtain a set of prediction scores. The higher the value of a particular prediction score, the more interesting and relevant the candidate is for follow-up.

The main motivation for developing the \texttt{SNGuess} method is the scientific use case of selecting young, extragalactic, and nearby SNe for additional follow-up observations.
Almost all of them become bright at the peak of their transient lifetime.
For our present analysis, we assume that the ZTF Bright Transient Survey (BTS; \citealt{fremling_zwicky_2020}, \citealt{perley_zwicky_2020}) has classified most of those transients to date, as this survey aims to classify all transients that at some point become brighter than 18.5 in magnitude.
We are interested in the feasibility of automatically detecting those transients even earlier in time. 
Therefore, we use labels from BTS to indicate relevance for follow-up of individual transients.

\begin{figure*}
\centering
\includegraphics[width=\textwidth]{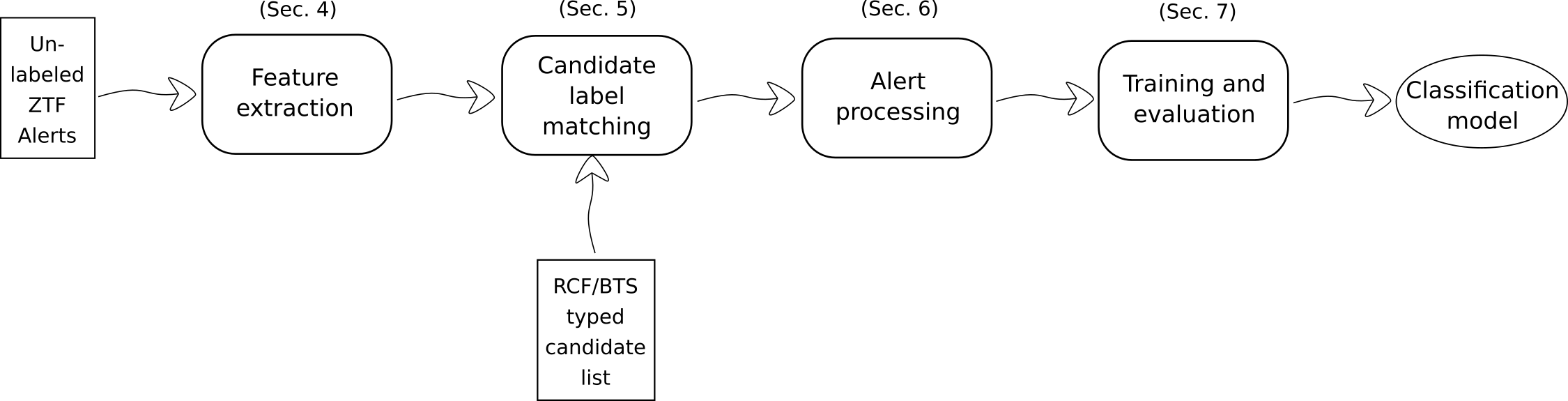}
\caption{Steps for \texttt{SNGuess} model generation. Each step in the process is described in a separate section, as indicated above the boxes.}
\label{fig:snguess_steps}
\end{figure*}

The process used to generate the core classification model of \texttt{SNGuess} can be separated into the following phases (see Fig. \ref{fig:snguess_steps}):
\vspace{-1ex}
\begin{enumerate}
    \item Alert pre-selection and feature extraction (Sect. \ref{sec:feature_extraction});
    \item Candidate label matching (Sect. \ref{sec:candidate_label});
    \item Alert processing (Sect. \ref{sec:alert_processing});
    \item Training and evaluation (Sect. \ref{sec:training}).
\end{enumerate}

In the
following sections, we discuss details of these steps together with the data sets used.

\section{Alert pre-selection and feature extraction}
\label{sec:feature_extraction}

\texttt{SNGuess} was trained and tested with data from the ZTF survey\footnote{In the future, ZTF alerts will contain forced photometry data. This study was made based only on alert photometry.}.
ZTF generated on average more than 150\,000 alerts per night for its public survey between May 2018 and July 2019. 
An initial pre-selection was made to only include alerts located outside the Galactic plane and with at least two detections and a \texttt{RealBogus} score larger than $0.3$.  \texttt{RealBogus} is an automatic classification system designed to identify observational artifacts.
Additionally, known Solar System objects flagged by the Minor Planet Center (MPC) are discarded.
This filtering process is similar to those typically performed by real-time observational surveys, and minimizes the impact from Solar System objects and stellar variability in the Galaxy.
The result of this filtering process is a set of 261\,417 astronomical alerts generated by the ZTF stream. This set contains photometric detections that took place between May 2018 and July 2019.

\begin{table*}
\caption{List of features used by \texttt{SNGuess}.}
\centering
\begin{tabular}{l c l c}
\hline\hline
\noalign{\smallskip}
Name & Type & Description & Source \\
\hline
\noalign{\smallskip}
tPredetect & Time & Time between final good upper limit and first detection. & Photometric \\
tLC & Time &  Duration (time between first and most recent detection). & Photometric \\
ndet & Int & Number of significant detections. & Photometric \\
peaked & Bool &  Is the lc estimated to be declining? & Photometric \\
pure & Bool & No significant nondetections after first detection. & Photometric \\
rising & Bool & Max brightness close to the most recent detection. & Photometric \\
norise & Bool & No (significant) detected rise. & Photometric \\
hasgaps & Bool & The light curve has a gap between detections of at least $30$ days. & Photometric \\
mPeak & Mag &  Magnitude at peak light (any band). Only calculated if peaked==True. & Photometric \\
mDet & Mag & Magnitude at first detection (any band). & Photometric \\
mLast & Mag & Magnitude of the current (i.e. latest) detection (any band). & Photometric \\
cPeak & $g$-$r$ &  Color at peak (if peaked and with $g$ and $r$). & Photometric \\
cDet & $g$-$r$ & Color at detection (if with $g$+$r$). & Photometric \\
cLast & $g$-$r$ & Color at last detection (if with $g$+$r$). & Photometric \\
slopeRise {g,r} & Mag/time & $g$ or $r$ mag slope between detection and peak (None if norise). & Photometric \\
slopeDecline {g,r} & Mag/time & $g$/$r$ magnitude slope between peak and last detection (None unless peaked). & Photometric \\
rb (med) & Float &  Median Real Bogus (all detections). & Photometric \\
drb (med) & Float & Median deep Real Bogus (if available, all detections). & Photometric \\
distnr & Pixel & Distance to nearest astronomical source in reference image & Metadata \\
magnr & Mag & Magnitude of nearest astronomical source in reference image. & Metadata \\
classtar & Float & Star/Galaxy classification score from SExtractor. & Metadata \\
sgscore1 & Float & Star/Galaxy score of closest astronomical source from PS1 catalog. & Metadata \\
distpsnr1 & Arcsec & Distance to closest astronomical source from PS1 catalog. & Metadata \\
sgscore2 & Float & Star/Galaxy score of next to closest astronomical source from PS1 catalog. & Metadata \\
distpsnr2 & Arcsec & Distance to next to closest astronomical source from PS1 catalog. & Metadata \\
neargaia & Arcsec & Distance to closest astronomical source from Gaia DR1. & Metadata \\
maggaia & Mag & Gaia (g-band) magnitude of closest astronomical source from Gaia DR1 catalog. & Metadata \\
\hline  
\end{tabular}
\tablefoot{The features are separated into \emph{photometric} (created based on the transient light curve provided in the alert) and \emph{metadata} (created based on the properties contained in the ZTF alerts of nearby detections in the references and in astronomical catalogues). While most of the features are self-explanatory, their explicit definitions can be found in Appendix \ref{app:feats}.}
\label{tab:risedec}
\end{table*}

We define a set of features (numerically measurable properties) to be extracted from the alert set that are relevant to our context. These are simple to calculate from alerts, regardless of whether they contain data for few or for many observations. Table \ref{tab:risedec} shows the full list of used features and their description.

The feature set can be divided in two subsets. The first one includes features that are related to the photometric observations contained in the alert. These are calculated using functions that depend only on the photometric points. Some examples of this kind of feature are the slope, color, duration, and peak magnitude of the light curves. The second subset contains the features that consist of values relating to the alert metadata or, more specifically, related to the transient candidate from which the photometric observations originate, such as its location in the sky, or its distance to the nearest astronomical source in a particular catalog in the PanStarrs \citep{kaiser_pan-starrs_2002} and GAIA \citep{perryman_gaia_2001} catalogs. Most of these numerical values are already present in the alert.

In general, the values of the alert metadata features tend to remain constant between alerts that refer to the same transient candidate. In constrast, the values of the photometric features usually differ between two alerts, depending on how the light curve evolves as more observations of the candidate are received.

The precision of photometric features varies strongly depending on the candidate luminosity. Even though the $5 \sigma$ detection limits of the ZTF is around 20 to 21 magnitudes, there can still be large uncertainties for detections that are fainter than 19.5. Therefore, it is not guaranteed that an alert with a detection of that magnitude will be distributed.
For this reason, any calculation made for these detections will have large uncertainties associated and problems with incompleteness.

Some of the features take Boolean values, such as the \texttt{peaked} field, which indicates whether the alert's light curve has a peak or not. Other features have integer or floating point values, such as the number of detections (\texttt{ndet}) or the distance to the nearest astronomical source (\texttt{distnr}).

The output of the feature calculation phase is one vector of feature values for each alert.
These feature vectors are arranged as matrix rows to form a feature matrix.

\section{Candidate label matching}
\label{sec:candidate_label}

The classification at the core of \texttt{SNGuess} is performed by a supervised model. However, none of the alerts that are generated by ZTF have information on the type of the observed candidate. Furthermore, no single metadata field in the alert can be directly used as a reliable indicator of follow-up relevance. This is why candidate type information has to be obtained from external data sources. Then, candidate relevance, and therefore a suitable relevance label, can be inferred from the candidate type, depending on what type of object should be followed up. In our case, these objects are young extragalactic SNe.

\begin{table}[]
    \caption{Main types assigned from BTS to the alert data set used for training \texttt{SNGuess}.}
    \centering
    \begin{tabular}{l c c}
        \hline \hline
        \textbf{Type} & \textbf{Num. of candidates} & \textbf{Num. of alerts} \\
        \hline
        SN Ia & 974 & 16942 \\
        SN II & 215 & 5022 \\
        AGN & 162 & 5573 \\
        CV & 58 & 693 \\
        SN IIn & 53 & 1476 \\
        SN IIP & 45 & 1790 \\
        SN Ic & 38 & 813 \\
        SN IIb & 34 & 611 \\
        SN Ia 91T-like & 31 & 854 \\
        SN Ib & 24 & 611 \\
        \hline
    \end{tabular}
    \label{tab:types}
\end{table}

Two external type lists are used for this purpose. First, a list of 4\,578 labeled candidates from the BTS survey is obtained, and these are then cross-matched with the candidates in the training set of ZTF alerts. 
The BTS classifications are mapped into ten general classes, shown in Table \ref{tab:types}.
This mapping is required as some labels have an interrogation mark suffixed, indicating that the type assigned to the candidate is uncertain. Some labels simply indicate that no type was found for the candidate (e.g., \texttt{None}, \texttt{unknown}). Other labels are very similar to each other or may refer to very specific subtypes (e.g., \texttt{SLSN-I} and \texttt{SLSN-I.5}, \texttt{SN Ib,} and \texttt{SN Ib/c}, etc). The primary goal of \texttt{SNGuess} is the detection of SN-like objects, and to do this at an early moment in their life cycle, when subclassification is highly unreliable. Furthermore, a training set with overly fine-grained labeling results in classes with very few examples.
Additionally, we use a second set with $495$ candidates typed as CV that were retrieved in March 2020 as a source of labels to perform the matching process described above.

We note that the training sample contains nearby SNe eventually classified by BTS (and therefore with positive labels), fainter but real astrophysical sources, and nonastronomical objects (both without labels).

\section{Alert processing}
\label{sec:alert_processing}

Once the features are extracted from the ZTF alerts and the candidate type information is obtained, we process the alerts based on their observations and metadata.
First, if a metadata field contains a measurement value that is nonphysical, it is replaced with a \texttt{null} value.

Alerts belonging to candidates with less than six "final" detections are removed from the training set. Most likely they would not have been part of any follow-up campaign (e.g., due to poor weather later), and thus risk biasing the training set. All alerts of candidates with sufficient data to eventually be classified (if sufficiently bright) are included individually.

In a subsequent step, alerts are separated into six different groups according to the number of their detections (\texttt{ndet}). For each one of these groups, we exclude those alerts that exceed a preset maximum duration between the first and most recent detection (\texttt{tPredetect}) and a duration between final good upper limit and first detection (\texttt{tLC}). These time ranges were chosen to be longer than the nominal ZTF northern sky survey cadence, meaning that an upper limit (i.e., nondetection) or previous detections should exist during normal operations. 
This additional filtering rejects detections made after a gap in the observations (e.g., due to visibility, scheduling issues, or weather). 
We perform this filtering because we are training \texttt{SNGuess} with mostly human-generated labels as a gold standard. We assume that an expert human agent is unable to systematically detect young transients without having a baseline to which its recent explosive behavior can be compared.
The corresponding groups and their maximum values are shown in Table \ref{table:CutVals}.

\begin{table}
    \caption{Alert groups by number of detections and their cut values.}
    \centering
    \begin{tabular}{l c c}
        \hline \hline
        \noalign{\smallskip}
        Num. of detections & \texttt{tPredetect} [JD] & \texttt{tLC} [JD] \\
        \hline
        \noalign{\smallskip}
        2 & $\leq 3.5$ & $\leq 3.5$ \\
        3 & $\leq 3.5$ & $\leq 6.5$ \\
        4 & $\leq 3.5$ & $\leq 6.5$ \\
        5 & $\leq 3.5$ & $\leq 10$ \\
        6 & $\leq 3.5$ & $\leq 10$ \\
        $[7,100]$ & $\leq 10$ & $\leq 90$ \\
        \hline
    \end{tabular}
    \tablefoot{A separate model is trained for each row.}
    \label{table:CutVals}
\end{table}

Finally, alerts that had a first detection with a magnitude below 16 are excluded from the training data set; these alerts were consistently found to be due to stellar variability and the same cut is applied in the live ZTF alert stream.
Initially, the training set contains 261\,417 alerts for 45\,765 different candidates. At the end of the data processing phase, the resulting set contains 109\,587 alerts for 35\,883 candidates.

\section{Training and evaluation}
\label{sec:training}

We use the \textit{XGBoost} \citep{chen_xgboost_2016} implementation of the Gradient Boosted Trees algorithm to obtain a decision tree ensemble model for the classification. Models trained by this algorithm have proven to be state of the art in classification, showing low out-of-sample errors in a variety of domains \citep{zhang_up--date_2017}. They can handle missing feature values out of the box, which simplifies the preliminary processing phase by making it unnecessary to replace them with generic values or to extrapolate them from existing data.

The Gradient Boosted Trees algorithm receives as input a series of hyper-parameters that control how the training takes place, and puts constraints on model parameters to keep the model from over-fitting to the training data. The first task of the learning phase is to explore different combinations of hyper-parameter values in order to estimate how well the models generated by the training algorithm perform and how this performance changes with each one of these combinations, before selecting the best hyper-parameter values.

\begin{table*}
\caption{Hyper-parameter space used for tuning the XGBoost model.}
\centering   
\begin{tabular}{l c c c c c c c c}
\hline \hline
\noalign{\smallskip}
\textbf{Hyper-parameter} & \textbf{Min. Value} & \textbf{Max. Value} & \textbf{2 det.} & \textbf{3 det.} & \textbf{4 det.} & \textbf{5 det.} & \textbf{6 det.} & \textbf{7 to 100 det.}\\
\hline
\noalign{\smallskip}
Column sample by level & 0.01 & 1.0 & 0.01 & 1.00 & 1.00 & 1.00 & 1.00 & 0.00 \\
Column sample by tree & 0.1 & 1.0 & 0.74 & 0.75 & 0.69 & 0.92 & 0.75 & 0.74 \\
Gamma & 0.0 & 40.0 & 0.14 & 1.65 & 3.81 & 2.49 & 1.65 & 0.13 \\
Learning rate & 0.0005 & 0.5 & 0.38 & 0.17 & 0.10 & 0.32 & 0.17 & 0.38 \\
Maximum depth & 2 & 12 & 12 & 9 & 3 & 11 & 9 & 12 \\
Minimum child node weight & 0.0001 & 0.5 & 0.32 & 0.04 & 0.42 & 0.21 & 0.04 & 0.32\\
Subsample & 0.01 & 1.0 & 0.79 & 0.48 & 0.60 & 0.36 & 0.48 & 0.79 \\
\hline
\end{tabular}
\tablefoot{The tuning process was done via a randomized search, and final optimal hyper-parameters obtained per model by number of detections. 200 points were randomly selected from the joint distribution of uniform probabilities within each hyper-parameter range. A five-fold cross validation was then performed for each of them and the mean performance was calculated. The point with the highest mean performance was selected as the optimum combination of hyper-parameter values.}
\label{tab:hyper}
\end{table*}

Numerical ranges for all of the hyper-parameters are defined in order to explore the space of possible hyper-parameter combinations (see Table \ref{tab:hyper}). Each one of the hyper-parameters is then assigned a uniform probabilistic distribution within its range. Next, a randomized search is performed in order to obtain an optimal hyper-parameter value combination. That is, $m$ points are randomly sampled from the joint distribution over all hyper-parameter variables. Each one of these points represents a combination of values for all hyper-parameters. For each one of them, training and evaluation of the model is performed with five-fold cross-validation. The mean and variance of the classification performance metric for all cross-validation folds are calculated before continuing with the next point sampled from the hyper-parameter joint distribution.
The model that is selected as the best is the one that obtains the highest performance metric across all its cross-validation folds.

This training and evaluation process is performed once for every group of alerts (alerts grouped by their number of detections; see Section \ref{sec:alert_processing}). 
For groups of alerts that have six or fewer detections, the following features are used as input for training: \texttt{mDet}, \texttt{mLast}, \texttt{tLC}, \texttt{rb}, \texttt{cDet}, \texttt{tPredetect}, \texttt{distnr}, \texttt{magnr}, \texttt{classtar}, \texttt{sgscore1}, \texttt{distpsnr1}, \texttt{neargaia}, and \texttt{maggaia}. 
For the final group, with alerts that have between 7 and 100 detections, all of the features indicated above are used as input for training, plus: \texttt{mPeak}, \texttt{cLast}, and \texttt{cPeak}.

After the training and evaluation phase, one set of boosted trees is produced for each one of the six groups into which the alerts were separated according to their number of detections. 
It is important to note that this division is done on an alert-by-alert basis, and not on an object-by-object basis. 
For instance, if an object has seven detections, one of its alerts will be considered in the two-detections group to start with, but later another one of its alerts will be considered in the 7-100 group (and its other alerts in the respective groups in between).
Table \ref{tab:hyper} shows the optimal hyper-parameters obtained after performing cross-validated training and evaluation over every alert group by number of detections.

\begin{figure}
\centering
\resizebox{\hsize}{!}{\includegraphics{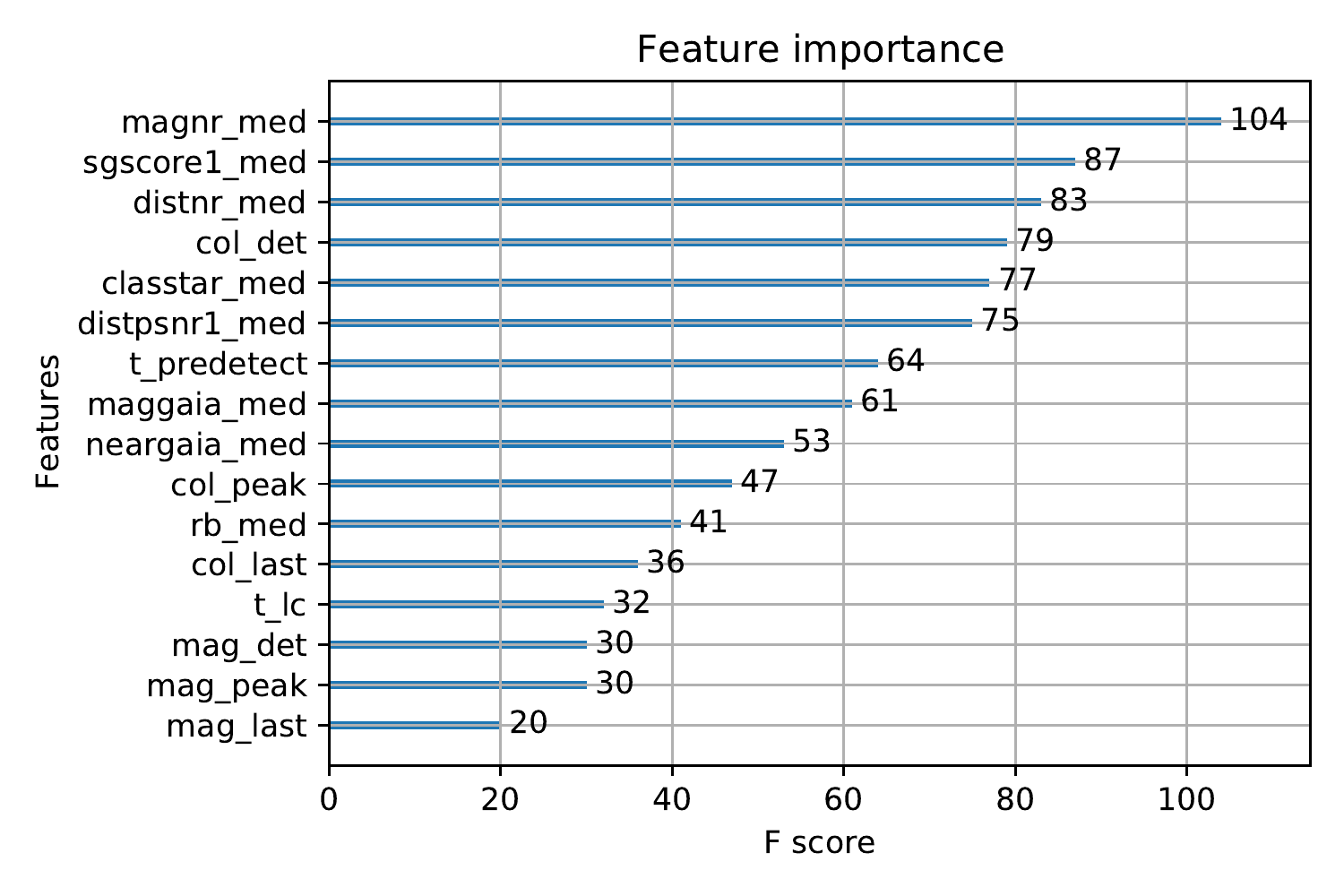}}
\caption{Ranking of most important features for classification with the 7 to 100 detections classifier, according to the F1-score metric.}
\label{fig:snguess_model_feat_importance}
\end{figure}

\emph{XGBoost} allows the user to easily generate a ranking of feature importance for the classification model. Figure \ref{fig:snguess_model_feat_importance} shows a list of the most important features for classification in the 7 to 100 detections classifier.

\section{Model deployment}
\label{sec:model_deployment}

\texttt{SNGuess} was designed as a light-weight system that is easy to both retrain and use. The data and procedures described so far have been made public in an \texttt{SNGuess} Git repository\footnote{\url{https://github.com/nmiranda/SNGuess}}. Users can access Python Jupyter notebooks with all of the steps in the data processing and training pipeline of \texttt{SNGuess}\footnote{\url{https://github.com/nmiranda/SNGuess/tree/main/notebooks}}. The results and plots of this article can be locally replicated, or the model retrained over a different set of features or labels.

\texttt{SNGuess} has been deployed as part of the \emph{AMPEL} real-time alert processing platform \citep{nordin_transient_2019}, and has been running on the servers of \emph{DESY} in Zeuthen, Germany, since April 2020.
Users can directly access the generated \texttt{SNGuess} scores of ZTF transients through the \texttt{AMPEL} API\footnote{\url{https://ampel.zeuthen.desy.de/api/live/docs}}.
Alternatively, users can install the core \texttt{AMPEL} packages and run the trained \texttt{SNGuess} model on raw ZTF light curves directly\footnote{\url{https://github.com/AmpelProject/Ampel-HU-astro/blob/main/notebooks/ampel_api_run_T2BrightSNProb.ipynb}}. 

Finally, \texttt{SNGuess} is used as one of the components of the selection process for young ZTF transients which are autonomously submitted to the Transient Name Server (TNS\footnote{\url{https://www.wis-tns.org/}}) under the sender name \texttt{AMPEL\_ZTF\_NEW}. The score assigned by \texttt{SNGuess} is a floating point value between $0$ and $1$, where a score closer to $1$ indicates that the candidate is more relevant for follow-up observations than one with a score closer to $0$.

\section{Testing}
\label{sec:testing}

We evaluated the performance of \texttt{SNGuess} over a set of 173\,402 alerts for 8\,969 candidates, received from October 8, 2020 to August 15, 2021.
These alerts remained after applying the same filter for poor-quality observations, as discussed in Section \ref{sec:alert_processing}.

Four classes of alerts were defined in order to generate performance metrics for \texttt{SNGuess}:

\begin{itemize}
    \item \textbf{Actual positive}: Any alert of a candidate eventually targeted by BTS;
    \item \textbf{Actual negative}: Any alert of a candidate not targeted by BTS;
    \item \textbf{Predicted positive}: Any alert with an \texttt{SNGuess} score above a certain threshold;
    \item \textbf{Predicted negative}: Any alert with an \texttt{SNGuess} score below or equal to the threshold.
\end{itemize}

The use of BTS labels as a measure of the "true" outcome assumes that this survey was $100\%$ efficient in terms of selecting transients by a human scanner. There are several reasons why this precision is not achieved in practice: observations of transients might have halted prior to reaching the BTS magnitude threshold due to weather or visibility, the transient might be intrinsically faint (or reddened), or it could be located in the core of a bright galaxy. It is therefore likely that the evaluation carried out here \emph{undervalues} the \texttt{SNGuess} performance.

\begin{table*}
\caption{Performance metrics for \texttt{SNGuess} over the test data set for different models, by number of detections of their alert training data set.}
\centering
\begin{tabular}{l c c c c c c c}
\hline\hline
\noalign{\smallskip}
\textbf{Metric} & \textbf{2 det.} & \textbf{3 det.} & \textbf{4 det.} & \textbf{5 det.} & \textbf{6 det.} & \textbf{7 to 100 det.} & \textbf{All models} \\
\hline
\noalign{\smallskip}
Precision (best=1.0, worst=0.0) & 0.473 & 0.563 & 0.642 & 0.675 & 0.671 & 0.962 & 0.876 \\
Recall (best=1.0, worst=0.0) & 0.537 & 0.781 & 0.827 & 0.877 & 0.898 & 0.830 & 0.821 \\
F1-score (best=1.0, worst=0.0) & 0.503 & 0.655 & 0.722 & 0.763 & 0.768 & 0.891 & 0.848 \\
MCC (perfect=+1.0, random=0.0, inverse=-1.0) & 0.398 & 0.539 & 0.596 & 0.641 & 0.617 & 0.787 & 0.728 \\
\hline
\end{tabular}
\label{tab:snguess_metrics}
\end{table*}

\begin{figure}
\centering
\resizebox{\hsize}{!}{\includegraphics{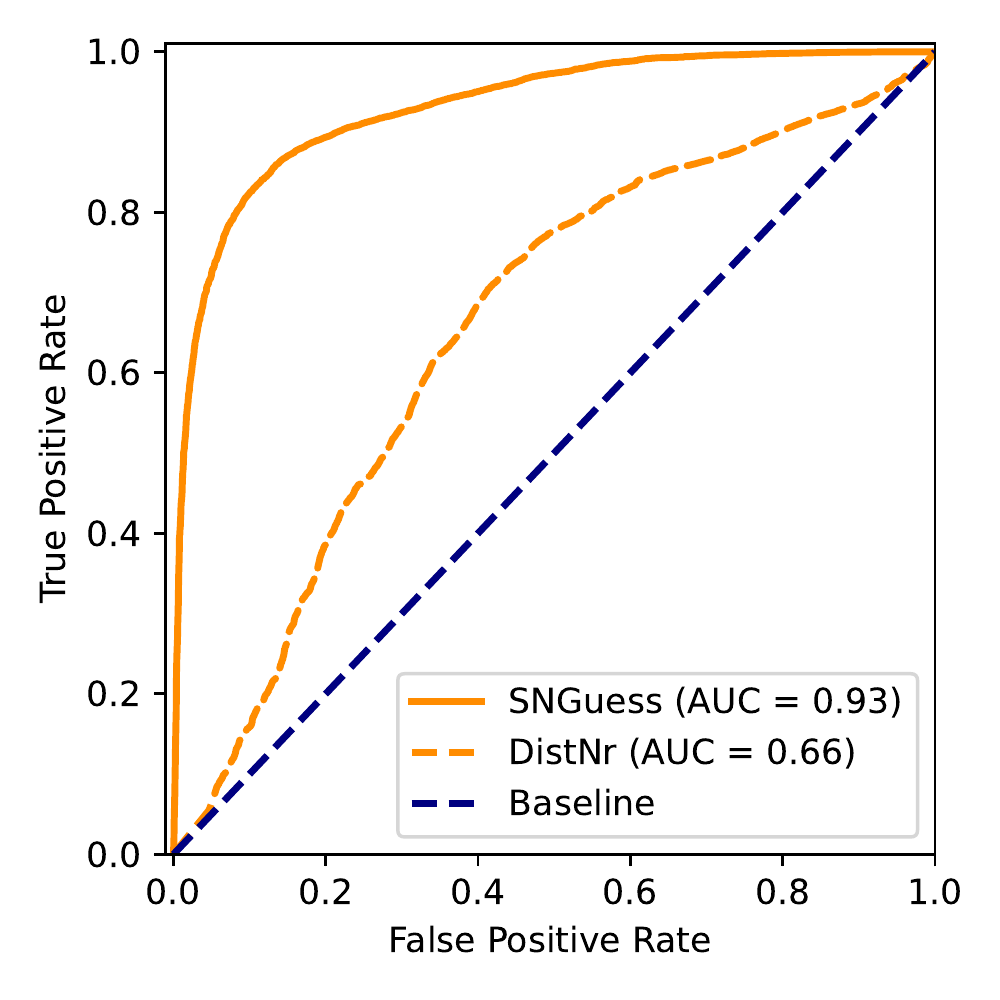}}
\caption{
ROC curve for \texttt{SNGuess} with the test data set (see Subsect. \ref{subsec:class_concepts}).
The area under the curve (ROC AUC) value of $0.93$ summarizes a good performance across different score thresholds for distinguishing between relevant and nonrelevant candidates. As a comparison point, we also see the ROC curve for performing a simple logistic regression classification with just the distance to nearest source as an independent variable. We can see that \texttt{SNGuess} shows a better performance than this simple classification (ROC AUC = 0.66).
}
\label{fig:snguess_hutns_class_roc}
\end{figure}

\begin{figure}
\centering
\resizebox{\hsize}{!}{\includegraphics{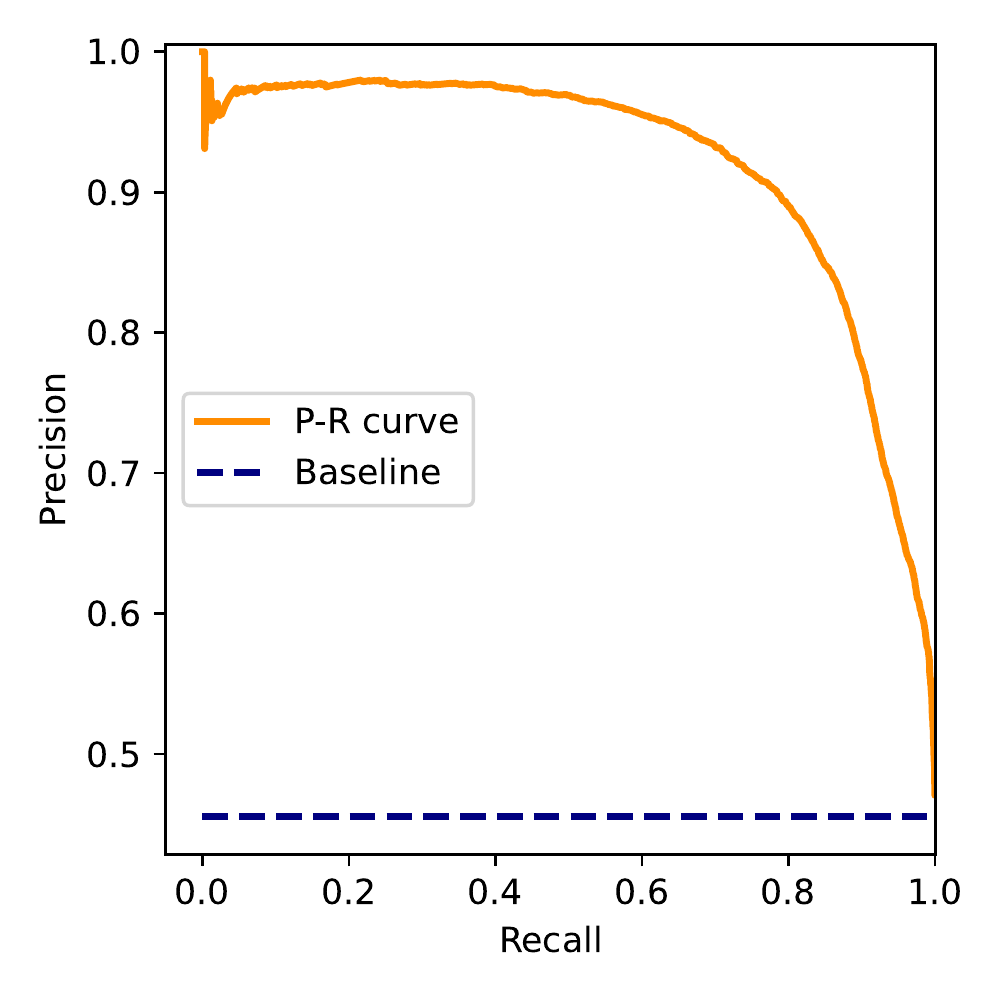}}
\caption{Precision--recall curve for selection by \texttt{SNGuess} from the test data set. The baseline performance shown by the blue dashed line corresponds to the ratio between relevant and total number of examples, and is an indicator of the imbalance between classes.}
\label{fig:snguess_hutns_class_prerec}
\end{figure}

Gradually increasing the \texttt{SNGuess} acceptance threshold while keeping record of the true-positive and false-positive rates produces the ROC curve shown in Fig. \ref{fig:snguess_hutns_class_roc}.
The ROC AUC value of $0.93$ reflects good performance over a variety of different transient ages and luminosities. 
We draw the same conclusion for the precision--recall curve of the selection, as shown in Fig. \ref{fig:snguess_hutns_class_prerec}. We realize that the curve is quite close to that of a perfect classification.

\begin{figure}
\centering
\resizebox{\hsize}{!}{\includegraphics{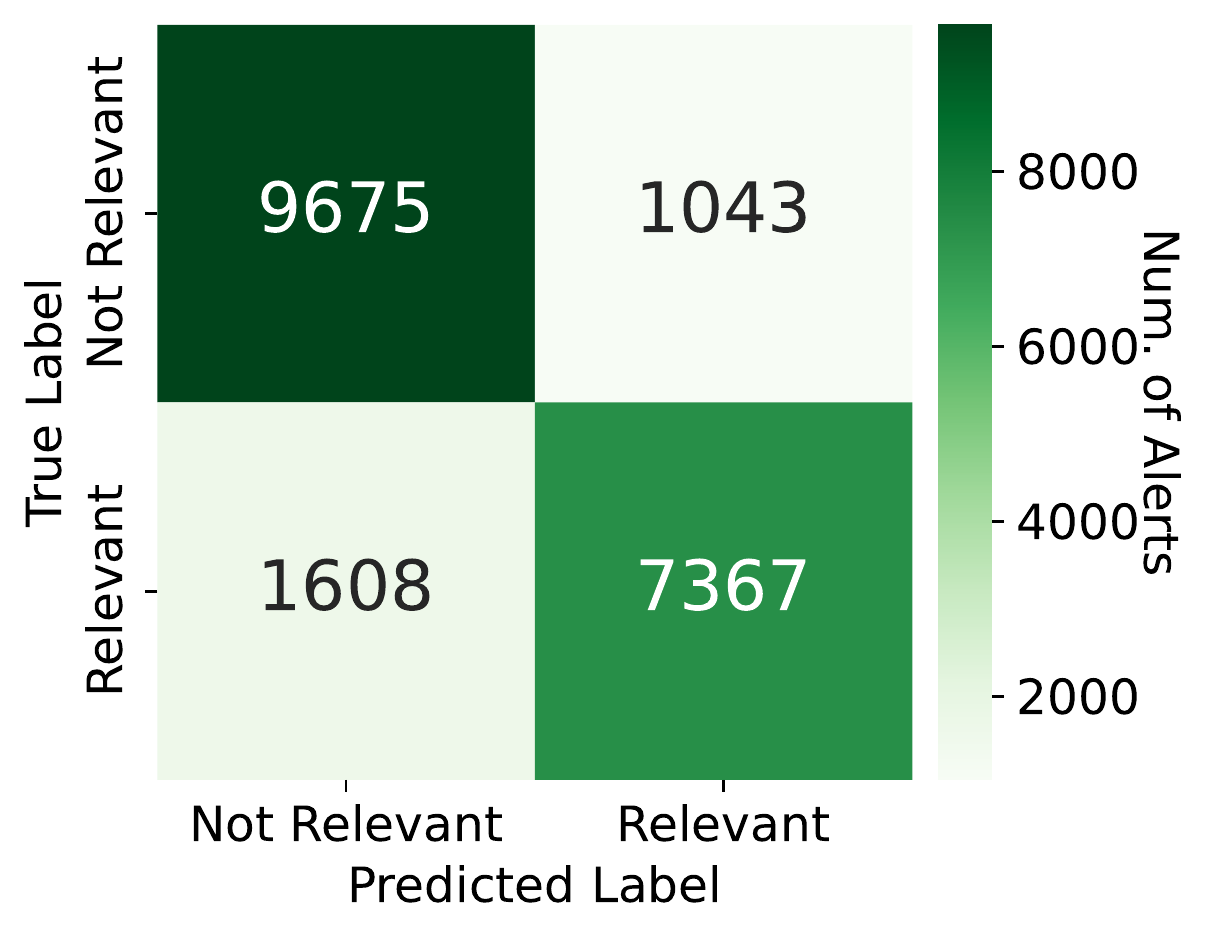}}
\caption{Confusion matrix of the selection by \texttt{SNGuess} from the test data set. The number in each box corresponds to the number of alerts fulfilling the \texttt{SNGuess} input criteria (see text for these).}
\label{fig:snguess_hutns_class_conf}
\end{figure}

If we fix the score threshold to $0.5$, and define all examples with a higher score as belonging to the positive selection or prediction class, we obtain a classification that produces the confusion matrix displayed in Fig. \ref{fig:snguess_hutns_class_conf}. We observe that false positives are less common than false negatives.
As we can see in Table \ref{tab:snguess_metrics}, \texttt{SNGuess} has a good performance in several metrics that are commonly used in binary classification tasks. Some of them, such as the F1-score and the Matthews correlation coefficient (MCC), are particularly important because of their robustness to class imbalance \citep{chicco_advantages_2020}.

\begin{figure}
\centering
\resizebox{\hsize}{!}{\includegraphics{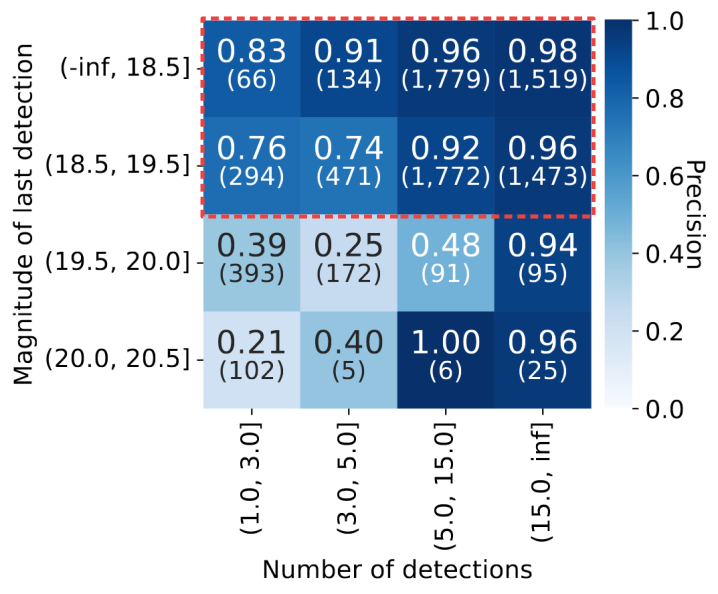}}
\caption{
Precision of the selection from alerts grouped by number of detections and magnitude of last detection for alerts fulfilling \texttt{SNGuess} run criteria.
The area highlighted in red, with last magnitude below $\sim 19.5$, correspond to alerts that are expected to have small photometric uncertainties. For fainter objects, ZTF photometry can exhibit significant uncertainties and variable detection efficiency.
}
\label{fig:snguess_hutns_prec}
\end{figure}

Astronomical use cases for a tool like \texttt{SNGuess} would typically include program limits in terms of the desired age or brightness of follow-up targets. Figure \ref{fig:snguess_hutns_prec} shows the precision of the selection by \texttt{SNGuess}  from alerts grouped by number of detections and magnitude of last detection. The number of detections works as an effective indicator of the potential age of the transients (which cannot be fully known at the time of the alert generation) as the nominal ZTF MSIP survey observes each field twice (g+r) every third day. The number of examples in each bin is displayed in parentheses in order to give an idea of the statistical significance. We again note that these are conservative numbers in that a fraction of the negative labels actually correspond to real extragalactic transients.

\texttt{SNGuess} consistently provides a precision above $90\%$ and approaching $100\%$ for transients suitable for follow-up with smaller scale facilities (magnitude $<19.5$) and of intermediate age ($5+$ detections). Our results show that \texttt{SNGuess} in this parameter range can, for example, efficiently reduce the amount of real-time scanning needed by human observers.

The numbers obtained in the lower two rows of the table in Fig. \ref{fig:snguess_hutns_prec} correspond to the area with large statistical uncertainties (see Sect. \ref{sec:alert_processing}), which will directly impact the feasibility of classification. 
In particular, due to the ZTF alert distribution mechanism, measurements in this brightness range might not be reported (even if eventually recognized by BTS).
While the results in Fig. \ref{fig:snguess_hutns_prec} are displayed in terms of the magnitude of last detection, the {mean} or typical magnitude of each alert is likely to be even fainter for the first two columns (as most real transients are increasing in brightness here), while it is likely to be brighter for the final column (where most transients have passed their peak). 

Science programs looking to target young extragalactic transients for immediate follow-up could make use of \texttt{SNGuess} to carry out automatic observations already at first detection for objects with magnitudes in bands g or r at $<19.5$ with reasonable precision ($\sim 75\%$). There is a drop in efficiency for young transients at larger magnitudes, and therefore \texttt{SNGuess} is more suitable here for use as an efficient first filter prior to a human decision regarding whether a transient should be targeted by the larger follow-up facilities. 

\begin{figure}
\centering
\resizebox{\hsize}{!}{\includegraphics{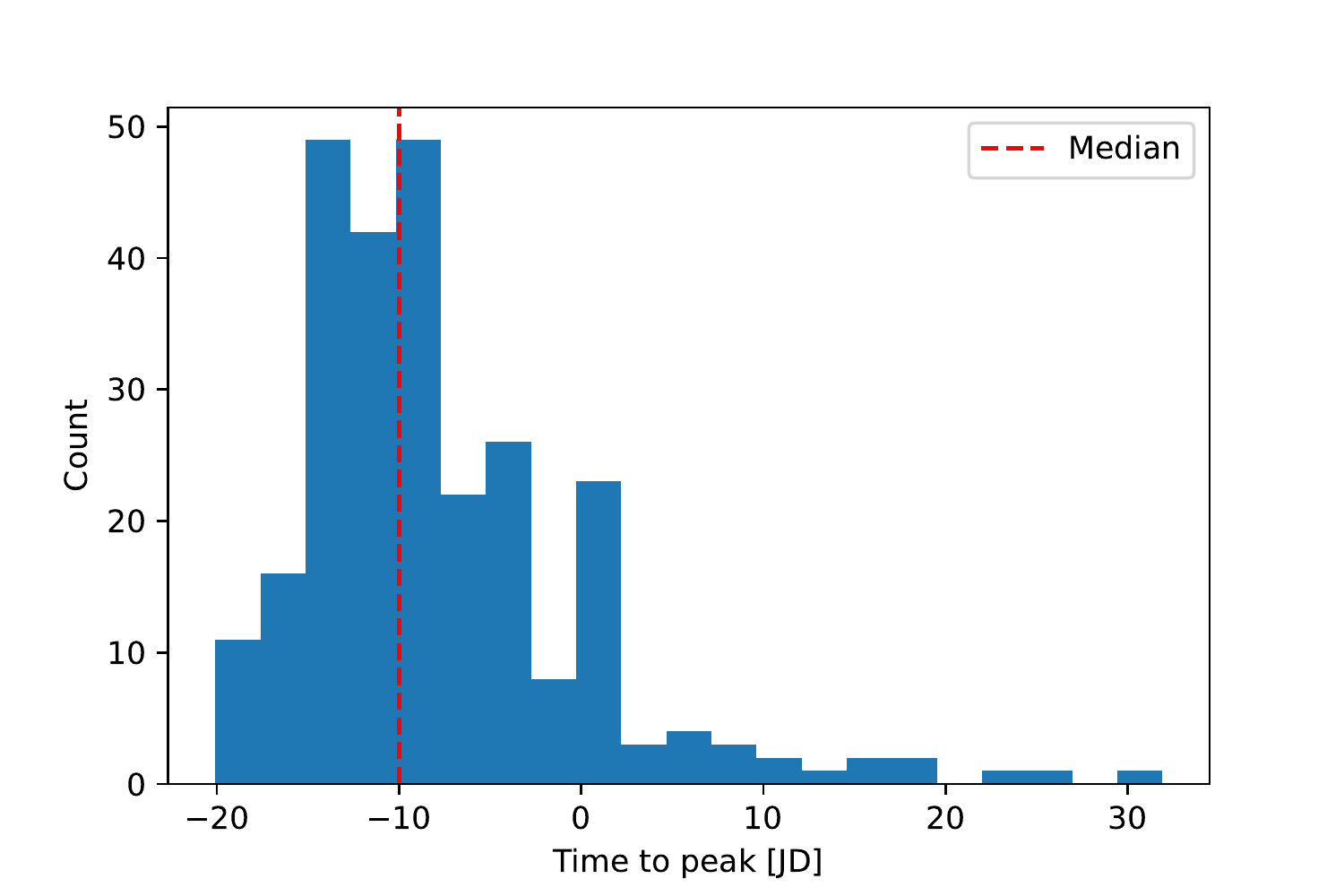}}
\caption{Distribution of SNeIa  candidate selection times with respect to light curve peak time. Two outlier points with extreme values were excluded from the plot: \texttt{ZTF21abbxtwv} and \texttt{ZTF18acydlux}. Both candidates show gaps of more than 30 days around peak magnitude in their light curves, which likely caused the peak-finding algorithm of BTS to give an incorrect detection time value. Similar observational irregularities are found for most SNe with a detection phase after peak.
}
\label{fig:snguess_hutns_phase}
\end{figure}

Finally, the number of significant ZTF {detections} that caused an {alert} to be generated is not identical to the real age of a transient, which is typically the focus of scientific interest. As SNeIa have a reasonably well-defined rise time to peak light ($18 \pm 2$ days), we can use the phase of candidate detection for transients later classified as SNeIa to estimate the time at which \texttt{SNGuess} would have selected a transient for follow-up. 
Figure \ref{fig:snguess_hutns_phase} shows the distribution of candidate-selection times with respect to the time of light curve peak. Candidate selection time is defined as the detection time (\texttt{jd\_last}) of the earliest alert for a particular candidate selected by \texttt{SNGuess}, and this is compared with the time of peak as catalogued by BTS. The median time of selection for SNIa candidates is close to $10$ days before peak. The early selection purity can easily be improved through, for example, matching to external catalogs of nearby galaxies, a step which was not done here in order to maintain the general applicability of the basic method.

\begin{figure}
\centering
\resizebox{\hsize}{!}{\includegraphics{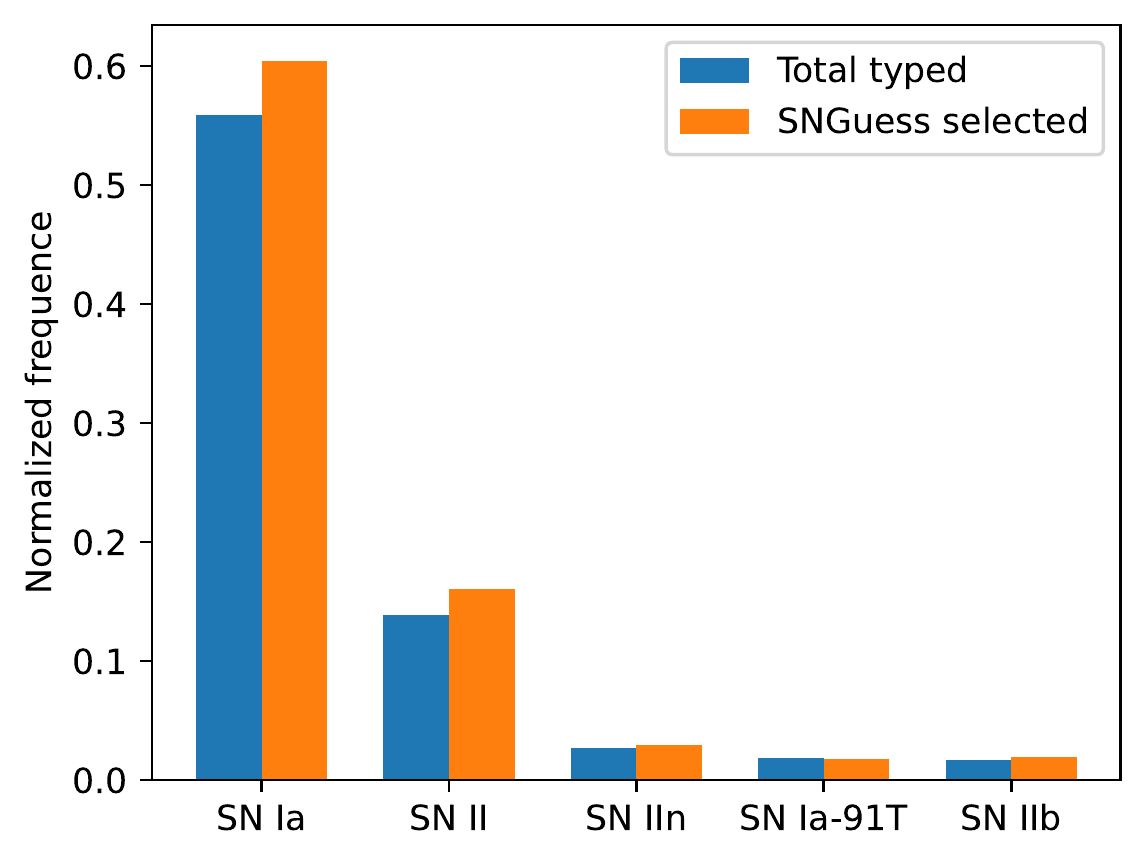}}
\caption{Comparison between the distribution of all typed candidates vs. SNGuess-selected candidates in the test data set; limited to the five most common types. The distributions are normalized in order to make it easier to compare relative biases.}
\label{fig:type_distribution}
\end{figure}

One should keep in mind that only a subset of the candidates that have been selected by BTS have also been assigned a type, and so by performing any analysis that involves confirmed types of candidates we are immediately subject to biases in the selection of BTS. 
In future work, we will enrich our data with further sources of labels, and also train and evaluate with data sets from other surveys, in order to have a more general assessment of the inherent type biases in the selection performed by \texttt{SNGuess}.

However, we can still compare the type distribution of all the candidates that have been classified with the type distribution of the candidates selected by \texttt{SNGuess} from the test data set. 
 Figure \ref{fig:type_distribution} shows this comparison for the five most common types. We see that even though the selection performed by \texttt{SNGuess} follows the overall distribution of the selection of BTS, the former is slightly more biased toward type Ia and type II SNe.

\section{Conclusions}
\label{sec:conclusions}

\texttt{SNGuess} is a light-weight machine-learning system designed to  assist in the selection of transients for potential follow-up observation from high-throughput alert streams generated by modern astronomical surveys. \texttt{SNGuess} was designed to work based on the alert content alone and without any external information. The output could therefore easily be further improved, for example, by taking into account redshift information and/or the rejection of galaxy core/AGN variability.

The average precision of \texttt{SNGuess} in terms of correctly predicting a candidate as being an interesting extragalactic transient is $88\%$, but this precision increases to $92\%$ to $98\%$ when examining transients brighter than $19.5$ mag with five or more detections.
This shows that \texttt{SNGuess} can be directly used for surveys looking to streamline the selection of interesting candidates for follow-up observations. 
The precision score obtained for transients in the same brightness class but with fewer detections is slightly lower, at $\sim 75\%$.
In addition, \texttt{SNGuess} is being used to search for faint, infant transients with few detections. However, the lower precision in this magnitude range ($\sim 30\%$) means that results need to be combined with visual inspection or catalog matching. Results in this magnitude range are affected by the increasing ZTF photometric uncertainties, which make both human and autonomous detection challenging.

As precision is a metric that is highly sensitive to class imbalance, we confirm that \texttt{SNGuess} produces good results according to other metrics, such as recall ($0.821$), F1-score ($0.848$), and MCC ($0.728$).

\texttt{SNGuess} was developed to help identify candidates of explosive transients  for spectroscopic follow-up as detected by 
the ZTF, but can easily be retrained to parse transients detected by the LSST at the Rubin Observatory. \textit{TiDES} is expected to be able to follow up all explosive transients detected by the LSST with magnitudes of $r_\mathrm{AB} \lesssim 22.5$ at peak \citep{swann_4most_2019}. While the spectroscopic sample obtained from this may require further augmentation or combination with other samples \citep{carrick_optimizing_2021}, an algorithm similar to \texttt{SNGuess} may be used to select appropriate candidates, particularly in the early part of the survey. 

Our evaluation procedure is limited, as our sets of alerts are not fully labeled (we do not have types for most of the candidates) and that the true labels are only obtained from the BTS survey.
Now that \texttt{SNGuess} is active we plan to carry out a dedicated follow-up of a subset of the \texttt{SNGuess} candidates in order to further evaluate, calibrate, and refine its results.

Our method may be improved by implementing a better strategy for hyper-parameter space search when training the classification model.
Furthermore, active learning could be incorporated in our pipeline. This would allow the use of human-in-the-loop and automatic follow-up results of transient observations as feedback for improving the quality of the classification. It could also allow us to retrain the model with more transient candidate sources for labeled data.
We would also like to explore different methods for estimating the performance of selection methods over partially labeled data sets by evaluating semi-supervised or unsupervised approaches.

The Python source code for \texttt{SNGuess} is fully available in the public repositories of \emph{AMPEL}, with additional data sets and notebooks that generate the results shown in this article\footnote{\url{https://github.com/nmiranda/SNGuess/tree/main/notebooks}}. The results of the live instance of \texttt{SNGuess} that is running on the servers at \emph{DESY} are freely accessible through the \emph{AMPEL} Open API\footnote{\url{https://ampel.zeuthen.desy.de/api/live/docs}}.
In addition, the repository contains an example notebook that shows how to obtain an \texttt{SNGuess} score for any ZTF candidate alert currently stored in the \emph{DESY} archives\footnotemark.
\footnotetext{Authentication for access to the \emph{DESY} archives is currently mediated via GitHub. If you don't have a GitHub user account with an active membership in the ZTF or \emph{AMPEL} organizations, please send an e-mail to \url{ampel-info@desy.de} in order to request access.}

\begin{acknowledgements}
We acknowledge the efforts of the BTS survey of the California Institute of Technology \citep{fremling_zwicky_2020, perley_zwicky_2020}, and of individual astronomers of the Humboldt-Universität zu Berlin in assigning types to transient candidates. Their results were instrumental to our supervised machine learning procedure. N. Miranda acknowledges the support of the Helmholtz Einstein International Berlin Research School in Data Science (HEIBRiDS).
\end{acknowledgements}

\bibliographystyle{aa}
\bibliography{aa}

\begin{appendix}

\section{Data inputs and mathematical formalization}
\label{app:Math}

\cite{chen_xgboost_2016} define a data set with $n$ examples and $m$ features $\mathcal{D} = \{(\mathbf{x}_i,y_i)\}, \ |\mathcal{D}| = n, \mathbf{x}_i \in \mathbb{R}^m, y_i \in \mathbb{R}$  as input for their Gradient Boosted Trees
learning model. We refer to $\mathbf{x}_i$, the \emph{input vector}, and $y_i$, the \emph{target value}.

When trying to set our photometric time-series data to comply with this format, we immediately face an issue: our data are composed of $l$ time series of variable length $\mathcal{T} = \{\mathbf{t}_j\}$, where $\mathbf{t}_j = (t, v)^{p_j}, \ t \in \mathbb{R}^+, \ v \in \mathbb{R}, \ p_j = |\mathbf{t}_j|, \ j \in [1,l], \ l \in \mathbb{N}^+_*$. We denote an astronomical time series, or \emph{light curve} with  $\mathbf{t}_j$. In the previous definition, $t$ is the \emph{time stamp} of a particular light curve measurement, and $v$ is the measurement value obtained at this time. Multiple $\mathbf{t}_j$ vectors can belong to the same astronomical candidate. This is due to the fact that photometric measurements are made for different band pass ranges, and measurements made in different bands do not necessarily have the same time stamp. In other words, a particular candidate has one light curve per band.

The approach we took to deal with this issue is to define a set of functions $\mathcal{F} = \{\mathit{f}_k\}, \ k \in [1,m]$, and then to apply this set of functions to all of the light curves for each candidate. This means that the $i$-th candidate will be assigned an input vector $\mathbf{x}_i = (\mathit{f}_1(\mathcal{T}_i), (\mathit{f}_2(\mathcal{T}_i), \dots (\mathit{f}_m(\mathcal{T}_i))$, where $\mathcal{T}_i$ is the set of light curves belonging to the $i$-th candidate. We refer to $\mathcal{F,}$ our set of \emph{feature functions}.

\section{Data inputs, data structure formalization}

There are several ways to store multivariate irregular-length time series as structured data. Let us suppose we have two candidates, for which we have measurements in at least one of $g$, $r$, and $i$ pass bands. If we want to put these measurements in a data table format, our first idea would be to assign each one of the candidates to a single row of the table. This we call a \emph{wide table} format. The result would then be of the form shown in Table \ref{table:WideTable}.

\begin{table*}
\caption{Wide table format.}
\label{table:WideTable}
\centering
\begin{tabular}{c c c c c c c c c c}
\textbf{id} & \textbf{g\_t1} & \dots & \textbf{g\_tm} & \textbf{r\_t1} & \dots & \textbf{r\_tm} & \textbf{i\_t1} & \dots & \textbf{i\_tm} \\
\hline
A & g\_t1(A) & \dots & g\_tm(A) & r\_t1(A) & \dots & r\_tm(A) & i\_t1(A) & \dots & i\_tm(A) \\
B & g\_t1(B) & \dots & g\_tm(B) & r\_t1(B) & \dots & r\_tm(B) & i\_t1(B) & \dots & i\_tm(B) \\
\end{tabular}
\end{table*}

In this case, $t_m$ would be the length of the longest light curve in our data set. The dimensions of this table, in general, would then be $n \times b \times t_m$, where $n$ is the number of candidates and $b$ is the number of pass bands. The advantages of this representation are its simplicity (all values are readily accessible) and the fact that it already has a format that can be accepted by most machine learning models (see Appendix \ref{app:Math}). In this case the table has one single obvious index: column \textbf{id}.

A wide table arrangement is useful when it is desirable to preserve the intuition present in general data science applications, that is, that each row of a table corresponds to an independent and identically distributed (i.i.d.) entity in the domain input.
The disadvantage of this representation is the sparsity of the table, as in most cases we have many fewer measurements in some pass bands than in others. In addition, the length of the light curves varies greatly between candidates. Because of this we are forced to fill the missing values with zeroes.

Another possible configuration would be what is called a \emph{nested table} configuration. In this arrangement, as in the wide table format, each row corresponds to a candidate. However, in this case the measurements are grouped in columns according to their type. Thus, we have one column for each band, and each value of the table contains the entire time series of the corresponding candidate at a certain band. Applying to our example case, the result would that displayed in Table \ref{table:NestedTable}.

\begin{table*}
\caption{Nested table format.}             
\label{table:NestedTable}      
\centering   
\begin{tabular}{c c c c}
\textbf{id} & \textbf{g} & \textbf{r} & \textbf{i} \\
\hline
A & g\_t1(A) \dots g\_tA(A) & r\_t1(A) \dots r\_tA(A) & i\_t1(A) \dots i\_tA(A) \\
B & g\_t1(B) \dots g\_tB(B) & r\_t1(B) \dots r\_tB(B) & i\_t1(B) \dots i\_tB(B) \\
\end{tabular}
\end{table*}

The advantage of this nested configuration is that the data are much less sparse, as a particular table value is empty only in cases when one of the candidates does not have any measurements for a given pass band. Furthermore, we have no constraints on the length of a given time series, as its full content is saved in a single cell of our table, and therefore we can easily store light curves of different lengths.

The disadvantage of this configuration is that we are no longer storing simple native data types in the table cells, but a composite series that has to be codified or serialized for storing its values, and there is no standard way of doing this. Also, we cannot perform mathematical operations with it directly, as it has no trivial matrix representation.

We can have a compromise between the two cases if we use what we call a \emph{semi-wide table} format. In this arrangement, there is a row for each combination of candidate and time of observation. Additionally, it has one column for each one of the observation bands. This way, each time stamp is associated with one or more observation values, depending on how many bands were observed at the very same instant as indicated by the time stamp. Applying to our example case, the result would be the one shown in Table \ref{table:SemiWideTable}.

\begin{table}
\caption{Semi-wide table format.}             
\label{table:SemiWideTable}      
\centering   
\begin{tabular}{c c c c c}
\textbf{id} & \textbf{time} & \textbf{g} & \textbf{r} & \textbf{i} \\
\hline
A & t1 & g(A, t1) & r(A, t1) & i(A, t1) \\
A & t2 & g(A, t2) & r(A, t2) & i(A, t2) \\
\dots \\
A & tA & g(A, tA) & r(A, tA) & i(A, tA) \\
B & t1 & g(B, t1) & r(B, t1) & i(B, t1) \\
B & t2 & g(B, t2) & r(B, t2) & i(B, t2) \\
\dots \\
B & tB & g(B, tB) & r(B, tB) & i(B, tB) \\
\end{tabular}
\end{table}

Again, the advantage of this format is that we have no constraint over time series length, and that we have no issues storing time series of different lengths. Also, it is very easy to append new values to a table in this format, as any new value is just a new row in the table, and the rows do not need to be in any particular order.

The disadvantage of this format is again sparsity, as there are almost no cases where for a particular instant in time there are measurements in more that one pass band. In other words, almost no observations of a particular candidate in different bands share exactly the same time stamp. Therefore, many rows of the table will have zeroes in at least two of its columns. Another disadvantage is that we no longer have a single column index. In this case we may consider a composite index made out of tuples of the columns \textbf{id} and \textbf{time}.

Yet another possible configuration is what we call \emph{long table} format. In this arrangement, a single row in the table corresponds to a single observation, and corresponds to a combination of candidate ID, time stamp, and band. Applying to our example, the result would be that shown in Fig. \ref{table:LongTable}.

\begin{table}
\caption{Long table format.}             
\label{table:LongTable}      
\centering   
\begin{tabular}{c c c c c}
\textbf{id} & \textbf{time} & \textbf{pass band} & \textbf{value} \\
\hline
A & t1 & g & g(A, t1) \\
A & t2 & g & g(A, t2) \\
A & t3 & r & r(A, t3) \\
\dots \\
A & tA & g & g(A, tA) \\
B & t1 & g & g(B, t1) \\
B & t2 & i & i(B, t2) \\
B & t3 & g & g(B, t3) \\
\dots \\
B & tB & r & r(B, tB) \\
\end{tabular}
\end{table}

The advantages of this configuration are the same as the previous one, with the difference that in this case the flexibility is even greater. We can arbitrarily append new measurements to the table as new rows, the rows do not need to be in any particular order, and we are not forced to fill with empty or zero values at all.

The disadvantage of this configuration is its length ($\mathcal{O}(n \times b \times t_m)$) and the fact that if we want to set an index to the table this will have to be composite of more than one column. If we can guarantee that for a single candidate there is not more than one measurement at a given time, then we can set \textbf{id} and \textbf{time} as index. If not, then the index will have to be composed of the columns \textbf{id}, \textbf{time}, and \textbf{pass band}.

Another common format to use for this type of data is what we call \emph{dictionary (or hashed index) format}. The idea is to have one single dictionary entry for each one of the candidates in our data set. The key of the entry is its \textbf{id}, and its value is a table with all the measurements for this candidate, as shown in Table \ref{table:DictTable}.

\begin{table}
\caption{Dictionary format.}             
\label{table:DictTable}
\centering   
\begin{tabular}{c c c c}
\textbf{dict key} & \textbf{dict value} & &  \\
\hline
id(A) & \textbf{time} & \textbf{pass band} & \textbf{value} \\
\hline
& t1 & g & g(A, t1) \\
& t2 & g & g(A, t2) \\
& \dots & \dots & \dots \\
& tA & g & g(A, tA) \\
\hline
id(B) & \textbf{time} & \textbf{pass band} & \textbf{value} \\
\hline
& tA & g & g(A, tA) \\
& t1 & g & g(B, t1) \\
& \dots & \dots & \dots \\
& tB & r & r(B, tB) \\
\end{tabular}
\end{table}

The advantages of this configuration are that, if we only (or most of the time) access by the candidate id, then it can be very efficient to work with this kind of structure. In this case, we retain the flexibility of the previous cases and we also have the ease of access of the wide table format.

The disadvantage in this case is that, again, we are in the same case as with the nested table, where the values of the dictionary entries may not necessarily of the same type of the dictionary; and thus we need to unpack the data structure of the value in order to access the individual measurements.
Finally, the data structure that we obtain once we apply our feature function set over our input data set has the format shown in Table \ref{table:FeatTable}.

\begin{table}
\caption{Feature table format.}
\label{table:FeatTable}
\centering
\begin{tabular}{c c c c c c}
\textbf{id} & \textbf{f\_1} & \textbf{f\_2} & \dots & \textbf{f\_m} & \textbf{target} \\
\hline
A & f\_1(A) & f\_2(A) & \dots & f\_1(A) & target(A) \\
B & f\_1(B) & f\_2(B) & \dots & f\_1(B) & target(B) \\
\end{tabular}
\end{table}

We have here a column for each one of our feature functions. This table may contain \texttt{null} values if the corresponding feature function cannot be applied to the  light curve of that candidate. The \textbf{target} column contains the target values for classification. This can be an integer number representing one of the possible classes, or a Boolean value in the case of a binary classification task. This data structure is already in a format that is accepted by most machine learning models, as it can be easily used as an input matrix and target vector.

For \texttt{SNGuess} we opted for internally using a dictionary format for representation of light curve data. This is very easy to derive from the AVRO format used by ZTF to distribute data via Apache Kafka streams. Then, after extracting features from the light curves, we use the feature table format to represent the light curves. All of the statistical analysis and machine learning model training, evaluation and testing is done over feature data in this format.

\section{Feature definition}
\label{app:feats}

\lstdefinestyle{customPythonStyle}{
  language=Python,
  breaklines=true
}
\lstset{basicstyle=\fontsize{5}{7}\selectfont\ttfamily,style=customPythonStyle}

\subsection*{\texttt{cut\_pp}}

Number of duplicate detections in the alert. By duplicate we mean one or more detections that have a time stamp value (floating point precision), in Julian date, that is the same as the time stamp of another detection in the alert. When a group of detections with the same time stamp is detected, the one with the highest real/bogus score is saved and the rest of them are labeled as duplicates and removed.

\subsection*{\texttt{jd\_det}}

Time stamp (floating point, Julian date) of the first (oldest) detection in the alert.

\subsection*{\texttt{jd\_last}}

Time stamp (floating point, Julian date) of the most recent detection in the alert.

\subsection*{\texttt{ndet}}

Total number of detections in the alert.

\subsection*{\texttt{mag\_det}}

Brightness, in apparent magnitude, of the first (oldest) detection in the alert.

\subsection*{\texttt{mag\_last}}

Brightness, in magnitude, of the most recent detection in the alert.

\subsection*{\texttt{t\_lc}}

Time span (in Julian date, floating point) between the oldest and the most recent detection in the alert.

\subsection*{\texttt{\{proptype\}\_med}}

\texttt{\{proptype\}} can be \texttt{rb}, \texttt{drb}, \texttt{distnr}, \texttt{magnr}, \texttt{classtar}, \texttt{sgscore1}, \texttt{distpsnr1}, \texttt{sgscore2}, \texttt{distpsnr2}, \texttt{neargaia}, or \texttt{maggaia}.

Median values of the properties of a particular detection  across all detections in the alert. These properties are mostly candidate-related and most of the time they do not vary between detections.

\subsection*{\texttt{bool\_pure}}

Indicates absence of upper limits after the first detection.

\subsection*{\texttt{t\_predetect}}

Time span between the latest upper limit prior to the first detection, and the first detection in the alert.

\subsection*{\texttt{bool\_peaked}}

Indicates if the light curve of the  alert has reached peak brightness.

\subsection*{\texttt{jd\_max}}

Time stamp of the peak brightness detection (when \texttt{bool\_peaked} is \texttt{true}).

\subsection*{\texttt{mag\_peak}}

Peak brightness, in magnitude, of the alert (when \texttt{bool\_peaked} is \texttt{true}).

\subsection*{\texttt{bool\_rising}}

Indicates whether the light curve of the  alert is rising in brightness.

\subsection*{\texttt{bool\_norise}}

Indicates whether there are no significant differences (within an error margin) between the peak magnitude of the alert and the detection magnitudes.

\subsection*{\texttt{bool\_hasgaps}}

Indicates whether there is a significant time gap (30 days) between consecutive detections in the alert.

\subsection*{\texttt{slope\_rise\_\{g,r\}}}

When \texttt{bool\_norise} is \texttt{false}, the slope of the rising part of the light curve of the  alert.

\subsection*{\texttt{slope\_fall\_\{g,r\}}}

Slope of the decline part of the light curve of the  alert, when this decline has taken place over less than 30 days.

\subsection*{\texttt{col\_\{det,last,peak\}}}

Color (magnitude difference between bands) at first, last, and peak detections.

\end{appendix}

\end{document}